\documentclass[conference, a4paper]{IEEEtran}
\IEEEoverridecommandlockouts
\usepackage{cite}
\usepackage{amsmath,amssymb,amsfonts}
\usepackage{algorithmic}
\usepackage{graphicx}
\usepackage{textcomp}
\usepackage[table,dvipsnames]{xcolor}
\usepackage[hyphens]{url}

\usepackage{multirow}
\usepackage{subfig}

\definecolor{light-gray}{gray}{0.8}
\definecolor{orange}{HTML}{ffddb3}
\definecolor{green}{HTML}{b3ffb3}
\definecolor{red}{HTML}{ffb3b3}
\definecolor{yellow}{HTML}{f8f8b9}

\def\BibTeX{{\rm B\kern-.05em{\sc i\kern-.025em b}\kern-.08em
    T\kern-.1667em\lower.7ex\hbox{E}\kern-.125emX}}

\begin{document}



\title{Commercial Evaluation of Zero-Skipping MAC Design for Bit Sparsity Exploitation in DL Inference}

\author{
\IEEEauthorblockN{
    Harideep Nair\IEEEauthorrefmark{1}\textsuperscript{\textsection},
    Prabhu Vellaisamy\IEEEauthorrefmark{1}\textsuperscript{\textsection},
    Tsung-Han Lin\IEEEauthorrefmark{2},
    Perry Wang\IEEEauthorrefmark{2},
   Shawn Blanton\IEEEauthorrefmark{1}, and
    John Paul Shen\IEEEauthorrefmark{1} 
 }
 
\IEEEauthorblockA{\IEEEauthorrefmark{1} Carnegie Mellon University} \IEEEauthorblockA{\IEEEauthorrefmark{2} MediaTek USA Inc.}}

\maketitle
\thispagestyle{empty}
\begingroup\renewcommand\thefootnote{\textsection}
\footnotetext{Equal contribution}
\pagestyle{empty}

\begin{abstract}

General Matrix Multiply (GEMM) units, consisting of multiply-accumulate (MAC) arrays, perform bulk of the computation in deep learning (DL).
Recent work has proposed a novel MAC design, Bit-Pragmatic (PRA), capable of dynamically exploiting bit sparsity.
This work presents \textit{Oz}MAC (\textit{Omit-zero}-MAC), a modified re-implementation of PRA, but extends beyond earlier works by performing rigorous post-synthesis evaluation against binary MAC design across multiple bitwidths and clock frequencies using TSMC N5 process node to assess commercial implementation potential.
We demonstrate the existence of high bit sparsity in eight pretrained INT8 DL workloads and show that 8-bit \textit{Oz}MAC improves all three metrics of area, power, and energy significantly by 21\%, 70\%, and 28\%, respectively.
Similar improvements are achieved when scaling data precisions (4, 8, 16 bits) and clock frequencies (0.5 GHz, 1 GHz, 1.5 GHz). For the 8-bit \textit{Oz}MAC, scaling its frequency to normalize the throughput, it still achieves 30\% improvement on both power and energy.
\end{abstract}

\begin{IEEEkeywords}
Zero-skipping multiply-accumulate, commercial TSMC N5 evaluation, bit sparsity, deep learning inference
\end{IEEEkeywords}

\section{Introduction and Background}
General matrix multiply (GEMM) hardware, employing large arrays of multiply-accumulate (MAC) units, is the core compute fabric for modern deep learning accelerators (DLAs) \cite{sze2020efficient}.
A conventional bit-parallel MAC unit consists of a combinational array multiplier and an adder to accumulate the product, and a register to store the value. Any improvement on the MAC unit design is replicated many fold in the large MAC arrays, yielding potential for significant reduction in hardware complexity of DLAs \cite{delmas2019bit, albericio2017bit, judd2016stripes, sharma2018bit}. Further, current industry standard for inference has moved from 32-bit floating-point (FP32) format to 16-bit floating-point (FP16) and 8-bit integer (INT8) formats. A recent study from IBM \cite{wang20188} summarizes the trend towards lower precision, highlighting imminent move towards 4-bit integer (INT4) and 2-bit integer (INT2) in the near future.

Recent work on Bit-Pragmatic (PRA) \cite{albericio2017bit} has proposed a novel MAC design that leverages bit sparsity (i.e., the number of `0' bits within a binary value) to perform bit serial compute efficiently by skipping over zero bits using simple serial shift-and-add compute.
In this work, we present a re-implementation of PRA with minor modifications for added hardware efficiency, called \textit{Oz}MAC, but the main contribution of this work is the rigorous state-of-the-art evaluation of \textit{Oz}MAC using commercial TSMC N5 (5nm) process node across multiple data precisions and clock frequencies, significantly extending beyond prior works utilizing TSMC 65nm technology and higher precision single-clock configurations.

Key contributions of our work are:

\begin{itemize}
    \item Present \textit{Oz}MAC based on Bit-Pragmatic (PRA), capable of exploiting dynamic bit sparsity by skipping over zero bits in binary values. This zero-skipping design in itself is not novel; the main focus of this work is its evaluation.
    \item Implement wide range of \textit{Oz}MAC designs using commercial design tools and TSMC N5 process design kit.
    \item Evaluate power-performance-area (PPA) for various data precisions (4-bits, 8-bits, 16-bits) and clock frequencies (500 MHz, 1 GHz, 1.5 GHz), against binary MAC.
    \item Demonstrate high bit sparsity in eight DL models leading to significant power reduction and how this can be used to increase \textit{Oz}MAC's throughput via frequency scaling.
\end{itemize}


The paper is organized as follows. \textit{Oz}MAC microarchitecture is briefly summarized in Section \ref{sec:microarch} followed by hardware evaluation methodology in Section \ref{sec:hweval}. We present sparsity and corresponding PPA evaluation in Section \ref{sec:sparsity}, followed by
bit-width and frequency scaling analysis in Sections \ref{sec:prec} and \ref{sec:freq} respectively. Finally, Section \ref{conclusion} presents key conclusions.

\section{OzMAC Microarchitecture and Design}
\label{sec:microarch}

\textit{Oz}MAC microarchitecture, derived from the Inner Product Unit within Bit-Pragmatic (PRA) \cite{albericio2017bit}, consists of three simple functional modules as shown in Fig. \ref{fig:ozmac_design}: 1) \textit{Oz-encoder}, 2) \textit{shifter}, and 3) \textit{accumulator}. \textit{Oz-encoder} is a Finite State Machine which keeps track of the current and next positions of `1' in the input bit pattern. Using this information, it outputs a one-hot encoded value capturing the bit positions of `1's every clock cycle for as many cycles as the number of `1's. For example, as illustrated in Fig. \ref{fig:ozmac_design}, the input `$0101_2$' is encoded as two one-hot values spanning two clock cycles: `$0100_2$' in the first cycle and `$0001_2$' in the next cycle. By doing this, it skipped over the two `0's and only incurs compute cycles for the `1's.
The \textit{Oz-encoded} input then goes to the shifter that determines the shift magnitude of the second input. The appropriately shifted second input is then added to the accumulator value.
The minor modification to PRA employed in \textit{Oz}MAC is the \textit{Oz}-encoder which feeds a 1-hot representation of the shift value to the shifter in contrast to PRA's oneffset generator which outputs a binary shift value. Employing 1-hot input simplifies the shifter hardware at the expense of more input lines (negligible overhead compared to reduction of shifter gate complexity).

\begin{figure}[t]
    \centering
    \includegraphics[width=0.65\columnwidth]{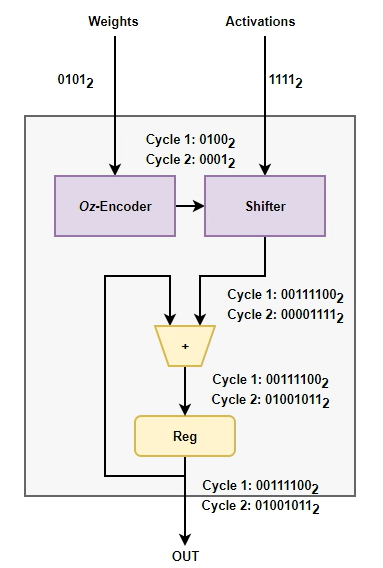}
    \caption{\textit{Oz}MAC (based on PRA \cite{albericio2017bit}) with example compute.}
    \label{fig:ozmac_design}
\end{figure}

\section{Hardware Framework for Evaluation}
\label{sec:hweval}

We perform rigorous, industry-standard evaluation of the \textit{Oz}MAC design to get accurate PPA and energy results and compare against a conventional bit-parallel bMAC. The technology library used for evaluation is the commercial TSMC N5 (5nm) process node, with Synopsys design tools employed for simulation, synthesis, and power calculations.

First, \textit{Oz}MAC RTL design is created in SystemVerilog, with functional verification performed using Synopsys VCS.
A synthetic dataset with 1000 sample weights and activation values is developed, with the values reflecting the sparsity levels of the DL benchmarks under consideration. This allows for the appropriate switching activity to be captured, as well as resultant average \textit{Oz}MAC compute cycles to be reported by the means of a testbench.
Next, lint check is performed on the SystemVerilog source files using Synopsys SpyGlass and then synthesis is performed to convert the RTL-level design into a gate-level netlist using Synopsys Design Compiler, sourcing TSMC N5 library files. Gate-level netlist simulation is then performed for verification and collection of the switching activity of the design in the form of a SAIF dump. The SAIF dump is then sourced along with the netlist to perform accurate power calculations using Synopsys PrimeTime PX.



\section{Sparsity Scaling Analysis}
\label{sec:sparsity}

Like PRA, \textit{Oz}MAC performs highly efficient shift-and-add operations and trades off latency for lower area and power. The ``omit-zero" capability of \textit{Oz}MAC is key to mitigating this latency overhead by exploiting dynamic bit sparsity in input data. In other words, higher bit sparsity (i.e., more zero bits in the input data) will result in shorter compute latency and thereby lower energy consumption.


\begin{table}[t]
\centering
\caption{Bit Sparsity and Cycle-Count Overhead for Pretrained Weights for Eight INT8 Quantized DL Benchmarks.}
\scalebox{1.05}{
\begingroup
\renewcommand{\arraystretch}{1}
 \begin{tabular}{|c|c|c|} 
 \hline
 DL & Average number of `1' bits & Bit Sparsity\\
 Benchmark & (Actual cycle-count overhead) & Percentage\\
 \hline
 \hline
 MobileNetV2 & 2.334 & 70.83\%\\
 \hline
 MobileNetV3 & 1.711 & 78.61\%\\
 \hline
 InceptionV3 & 2.430 & 69.62\%\\
 \hline
 ShuffleNetV2 & 2.583 & 67.71\%\\
 \hline
 GoogleNet & 2.461 & 69.24\%\\
 \hline
 ResNet18 & 2.398& 70.02\%\\
 \hline
 ResNet50 & 2.495 & 68.81\%\\
 \hline
 ResNeXt101 & 2.289 & 71.39\%\\
 \hline
 \end{tabular}
 \endgroup
 }
  \label{tab:dnn_sparsity}
\end{table}


As illustrated in Table \ref{tab:dnn_sparsity}, we use eight pretrained and quantized INT8 models, available as part of PyTorch's Torchvision library, that are widely used in state-of-the-art DL literature.
Layer-by-layer analysis of the converged weights and activation values resulting from running ImageNet benchmark inputs illustrate the sparsities inherent in these benchmarks. For each model, we extract the average number of `0' bits in every 8-bit weight value across all the layers and calculate the bit sparsity as the percentage of `0' bits over the total number of bits. The DL models have close to 70\% bit sparsity with MobileNetV3 having the highest sparsity of 78.61\%. Table \ref{tab:dnn_sparsity} also shows the number of `1' bits, which is equal to the compute latency in cycles. It can be seen that the effective compute latency for 8-bit \textit{Oz}MAC owing to bit sparsity is between 1.7-2.5 cycles, much lower than the worst-case latency of 8 cycles. Comparing this to 1 cycle latency of bMAC, \textit{Oz}MAC's power consumption must be 1.7-2.5x lower than that of bMAC to achieve similar energy efficiency.




\begin{table}[t]
\centering
\caption{TSMC N5 PPA and Energy (averaged across eight DL benchmarks) for 8-bit OzMAC and bMAC at 500 MHz.}
\scalebox{1}{
\begingroup
\renewcommand{\arraystretch}{1}
 \begin{tabular}{|c|c|c|c|c|} 
 \hline
 MAC & Area & Power & Latency & Energy\\
 Hardware & ($\mu$m\textsuperscript{2}) & (mW) & (ns) & (pJ)\\
 \hline
 \hline
 \rowcolor{red} bMAC & 25.361 & 0.084 & 2 & 0.167  \\
 \hline
 \rowcolor{green} OzMAC & 19.996 & 0.025 & 4.76 & 0.120\\
 \hline
 \rowcolor{yellow}\% Improvement & 21.2 & 69.7 & - & 28.0\\
 \hline
 \end{tabular}
 \endgroup
 }
  \label{tab:8bit}
\end{table}

\begin{figure}[t]
    \centering
    \includegraphics[width=0.8\columnwidth]{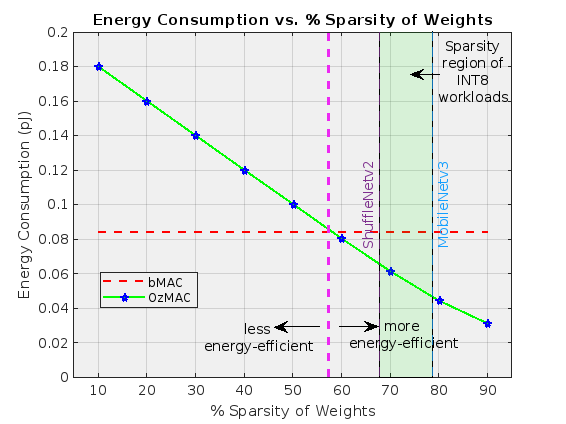}
    \caption{Energy consumption vs. \% bit-sparsity. Green-shaded region depicts the sparsity regions for Table \ref{tab:dnn_sparsity} workloads.}
    \label{fig:energy_sparsity}
\end{figure}


Table \ref{tab:8bit} provides the die area, power, latency and energy consumption of \textit{Oz}MAC and bMAC averaged across the eight DL benchmark models. Note that the power consumption values are obtained via PTPX using benchmark-specific test vectors that capture bit sparsity characteristics. The operating frequency for both designs is 500~MHz. An 8-bit conventional bMAC computes 1 MAC operation in 2~ns (1 cycle) while consuming about 25~$\mu$m\textsuperscript{2} area, 84~$\mu$W power, and 167~fJ energy, whereas \textit{Oz}MAC only consumes about 20~$\mu$m\textsuperscript{2} area, 25~$\mu$W power, and 120~fJ energy while incurring 4.76~ns latency on average. Compared to conventional bMAC, this amounts to 21\% less die area, 70\% less power, and 28\% less energy with 2.38x higher latency. This significant improvement in all three metrics can be attributed to three key factors: 1) simpler shift-and-add hardware with less area and leakage power footprint, 2) serial \textit{Oz}-encoder that enables significant reduction in signal transitions at the input stage, thereby improving dynamic power, and 3) capability to exploit the high bit sparsity present in the DL benchmarks (Table \ref{tab:dnn_sparsity}).

Given the power consumption values from Table \ref{tab:8bit}, it can be seen that \textit{Oz}MAC reduces power by 3.36x on average. This implies that for an 8-bit \textit{Oz}MAC design, it can incur up to $3.36$ clock cycles on latency overhead per MAC operation, before its energy consumption exceeds that of bMAC. 
We can calculate the minimum bit sparsity needed for \textit{Oz}MAC to maintain superior energy efficiency as $1-\frac{3.36}{8} = 58\%$. Fig. \ref{fig:energy_sparsity} plots the energy consumption across varying bit sparsity, to demonstrate this cross-over point at 58\% sparsity. Interestingly, all eight DL benchmarks exhibit bit sparsity higher than the threshold of 58\% as can be seen from Fig. \ref{fig:energy_sparsity}. Significant reduction in power consumption, coupled with sparsity-induced latency reduction, allows \textit{Oz}MAC to maintain superior energy efficiency over bMAC in spite of multi-cycle latency overhead. For throughput-sensitive applications, the higher latency of \textit{Oz}MAC can be addressed via frequency scaling, as will be demonstrated later in Section \ref{sec:iso_throughput}.


\textbf{Key Takeaway:} \textit{Oz}MAC achieves significant reduction in area, power and energy relative to bMAC for typical DL workloads, by exploiting inherent bit sparsity.


\section{Precision Scaling Analysis}
\label{sec:prec}


Inference precision for DL workloads has been trending from 16-bits in the past to the current 8-bits with projection further down to 4-bits.
%
%
%
%
Table \ref{tab:prec} provides TSMC N5 PPA for five integer precision configurations: 1) 4-bit weights, 4-bit activations (4x4), 2) 4-bit weights, 8-bit activations (4x8), 3) 8-bit weights, 8-bit activations (8x8), 4) 8-bit weights, 16-bit activations (8x16), and 5) 16-bit weights, 16-bit activations (16x16). The mixed precisions, 4x8 and 8x16, are used to accommodate typical workloads that demand higher activation precision compared to weight precision.
The corresponding area and power results are also plotted in Fig. \ref{fig:area-power}.

\begin{table}[t]
\centering
\caption{TSMC N5 PPA at 500~MHz across varying bit precision of weights and activations: 4 bits, 8 bits and 16 bits.}
\scalebox{1}{
\begingroup
\renewcommand{\arraystretch}{0.95}
 \begin{tabular}{|c|c|c|c|c|} 
 \hline
 MAC Hardware & Area & Power & Latency & Energy\\
 (wgt x act) & ($\mu$m\textsuperscript{2}) & (mW) & (ns) & (pJ)\\
 \hline
 \hline
 \rowcolor{red} bMAC (4x4) & 5.451 & 0.015 & 2     & 0.031\\
 \hline
 \rowcolor{green} OzMAC (4x4)    & 4.712 & 0.008 & 2.794 & 0.022\\
 \hline
 \rowcolor{yellow}\% Improvement   & 13.6  & 49.4  & -     & 29.2\\
 \hline
 \hline
 \rowcolor{red} bMAC (4x8) & 9.693 & 0.031 & 2 & 0.061  \\
 \hline
 \rowcolor{green} OzMAC (4x8) & 8.3752 & 0.013 & 2.794 & 0.035\\
 \hline
 \rowcolor{yellow}\% Improvement & 13.6 & 58.5 & - & 42.0\\
 \hline
 \hline
 \rowcolor{red} bMAC (8x8) & 25.361 & 0.084 & 2 & 0.167  \\
 \hline
 \rowcolor{green} OzMAC (8x8) & 19.996 & 0.025 & 4.76 & 0.120\\
 \hline
 \rowcolor{yellow}\% Improvement & 21.2 & 69.7 & - & 28.0\\
 \hline
 \hline
 \rowcolor{red} bMAC (8x16) & 45.282 & 0.177 & 2 & 0.355  \\
 \hline
 \rowcolor{green} OzMAC (8x16) & 30.909 & 0.041 & 4.76 & 0.196\\
 \hline
 \rowcolor{yellow}\% Improvement & 31.7 & 76.8 & - & 44.9\\
 \hline
 \hline
 \rowcolor{red} bMAC (16x16) & 74.199 & 0.297 & 2 & 0.594 \\
 \hline
 \rowcolor{green} OzMAC (16x16) & 60.608 & 0.065 & 9.28 & 0.601\\
 \hline
 \rowcolor{yellow}\% Improvement & 18.3 & 78.2 & - & -1.2\\
 \hline
 \end{tabular}
 \endgroup
 }
  \label{tab:prec}
\end{table}


\begin{figure}[t]
  \centering
  \subfloat[Die Area]{\includegraphics[width=0.24\textwidth]{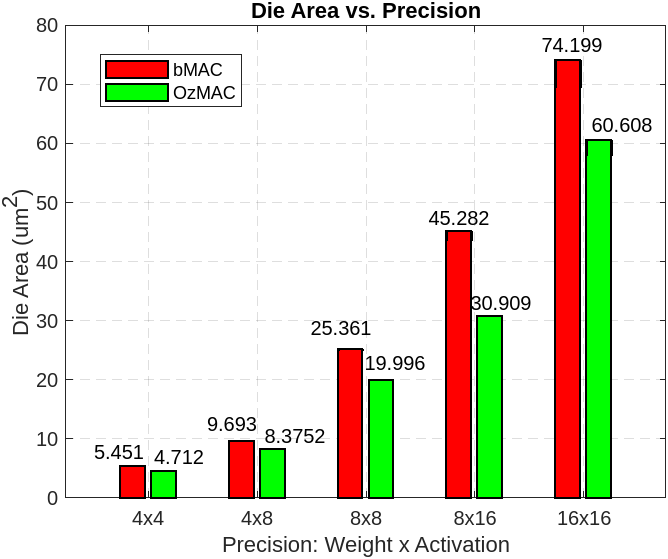}\label{fig:area}}
  \hfill
  \subfloat[Power]{\includegraphics[width=0.24\textwidth]{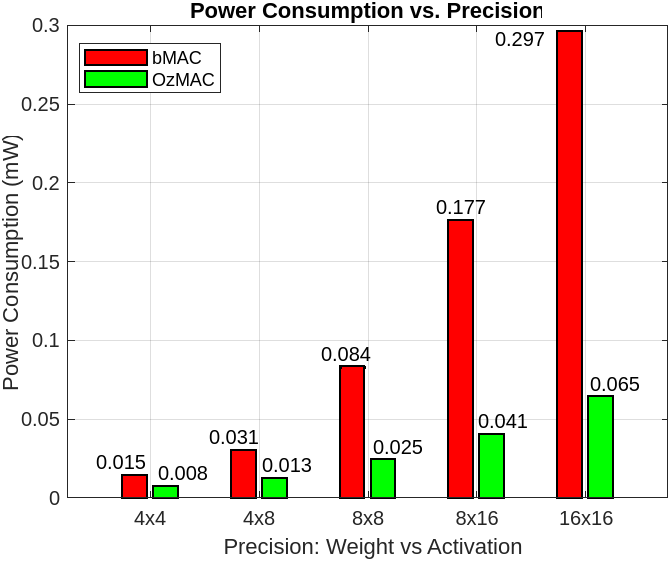}\label{fig:power}}
  \hfill
  \caption{Die area and power costs vs precision configurations.}
  \label{fig:area-power}
\end{figure}



Based on Table \ref{tab:prec}, the smallest (4x4) \textit{Oz}MAC and bMAC designs consume 4.7~$\mu$m\textsuperscript{2} area, 8~$\mu$W power, 22~fJ energy, and 5.4~$\mu$m\textsuperscript{2} area, 15~$\mu$W power, 31~fJ energy, respectively.
Compared to 4x4 designs, the largest (16x16) \textit{Oz}MAC incurs about 13x, 8x and 27x increase whereas 16x16 bMAC incurs close to 14x, 20x and 20x increase in area, power and energy, respectively. Both 16x16 designs yield comparable energy while \textit{Oz}MAC still possesses area and power benefits. This indicates going beyond 16-bits for \textit{Oz}MAC is not beneficial.

From Fig. \ref{fig:area-power}, area for \textit{Oz}MAC and bMAC scale up in a similar fashion almost linearly with respect to product of weight and activation bits. However, \textit{Oz}MAC's power consumption scales much better than that of bMAC, which incurs a much sharper increase with precision. Relative to bMAC, 8x16 \textit{Oz}MAC delivers the most area benefit (32\% improvement), whereas the mixed precision 4x8 and 8x16 \textit{Oz}MAC designs offer the highest energy benefit up to 45\%. Mixed precision designs deliver the highest energy improvements, as they can leverage the lower of the two precisions for \textit{Oz}-encoding, incurring minimum latency (and thereby energy) overhead while taking advantage of the lower hardware complexity.
Power benefits increase monotonically with precision due to the serial nature of \textit{Oz} computation with signal transitions that get relatively sparser with higher precision.

\textbf{Key Takeaway:} \textit{Oz}MAC is more area and power-efficient than bMAC across all precision configurations, and more energy efficient across all but one (16x16) configuration. Energy consumption for both designs evens out at 16-bit weight precision, beyond which \textit{Oz}MAC becomes inefficient due to high latency overhead.

\section{Frequency Scaling Analysis}
\label{sec:freq}

In this section, we evaluate two types of frequency scaling to assess the effects on PPA trends between \textit{Oz}MAC and bMAC.

\begin{table}[h]
\centering
\caption{TSMC N5 PPA for INT8 (8-bits) OzMAC across varying frequencies: 500~MHz, 1~GHz and 1.5~GHz.}
\scalebox{1}{
\begingroup
\renewcommand{\arraystretch}{1}
 \begin{tabular}{|c|c|c|c|} 
 \hline
 MAC & Power & Latency & Energy\\
 Hardware & (mW) & (ns) & (pJ)\\
 \hline
 \hline
 \rowcolor{red} bMAC (0.5~GHz) & 0.084 & 2 & 0.167  \\
 \hline
 \rowcolor{green} OzMAC (0.5~GHz) & 0.025 & 4.76 & 0.120\\
 \hline
 \rowcolor{yellow}\% Improvement & 69.7 & - & 28.0\\
 \hline
 \hline
 \rowcolor{red} bMAC (1~GHz) & 0.166 & 1 & 0.166  \\
 \hline
 \rowcolor{green} OzMAC (1~GHz) & 0.050 & 2.38 & 0.118\\
 \hline
 \rowcolor{yellow}\% Improvement & 70.1 & - & 28.7\\
 \hline
 \hline
 \rowcolor{red} bMAC (1.5~GHz) & 0.251 & 0.667 & 0.167 \\
 \hline
 \rowcolor{green} OzMAC (1.5~GHz) & 0.075 & 1.587 & 0.119\\
 \hline
 \rowcolor{yellow}\% Improvement & 70.2 & - & 29.0\\
 \hline
 \end{tabular}
 \endgroup
 }
  \label{tab:freq1}
\end{table}


\subsection{Iso-Frequency Evaluation}
\label{sec:iso_freq}

As can be seen from Table \ref{tab:freq1}, \textit{Oz}MAC consumes 50~$\mu$W power and 118~fJ energy at 1~GHz, and only 75~$\mu$W and 119~fJ even at 1.5~GHz.
At all three frequencies, \textit{Oz}MAC improves power and energy by almost 70\% and 29\% respectively.
As expected, power consumption scales linearly with frequency and energy stays almost constant since power increases and latency (due to clock period) reduces by similar amounts. 

\begin{table}[h]
\centering
\caption{TSMC N5 PPA for OzMAC and bMAC across varying bit precisions at throughput-matching frequencies.}
\scalebox{1}{
\begingroup
\renewcommand{\arraystretch}{1}
 \begin{tabular}{|c|c|c|c|c|} 
 \hline
 MAC Hardware & Freq & Power & Latency & Energy\\
 (wgt x act) & GHz & (mW) & (ns) & (pJ)\\
 \hline
 \hline
 \rowcolor{red} bMAC (4x4) & 0.5 & 0.015 & 2 & 0.031  \\
 \hline
 \rowcolor{green} OzMAC (4x4) & 0.7 & 0.011 & 2 & 0.022 \\
 \hline
 \rowcolor{yellow}\% Improvement & - & 29.2 & Equal & 29.3 \\
 \hline
 \hline
 \rowcolor{red} bMAC (4x8) & 0.5 & 0.031 & 2 & 0.061  \\
 \hline
 \rowcolor{green} OzMAC (4x8) & 0.7 & 0.018 & 2 & 0.036\\
 \hline
 \rowcolor{yellow}\% Improvement & - & 41.5 & Equal & 41.6\\
 \hline
 \hline
 \rowcolor{red} bMAC (8x8) & 0.5 & 0.084 & 2 & 0.167  \\
 \hline
 \rowcolor{green} OzMAC (8x8) & 1.2 & 0.059 & 2 & 0.118\\
 \hline
 \rowcolor{yellow}\% Improvement & - & 29.5 & Equal & 29.6\\
 \hline
 \hline
 \rowcolor{red} bMAC (8x16) & 0.5 & 0.177 & 2 & 0.355  \\
 \hline
 \rowcolor{green} OzMAC (8x16) & 1.2 & 0.096 & 2 & 0.192\\
 \hline
 \rowcolor{yellow}\% Improvement & - & 46.0 & Equal & 46.0\\
 \hline
 \end{tabular}
 \endgroup
 }
  \label{tab:freq2}
\end{table}

\subsection{Iso-Latency Evaluation}
\label{sec:iso_throughput}
\textit{Oz}MAC's area-power-energy improvements are achieved at the cost of increased latency (1.4x for 4 bits and 2.4x for 8 bits). Such \textit{Oz}MAC designs are ideal for edge inference applications that can tolerate the slight increase in latency (and reduction in throughput) but with stringent area/power/energy constraints. Here, we show that \textit{Oz}MAC can even be used effectively for higher throughput with higher clock frequency.

To bridge the latency gap between \textit{Oz}MAC and bMAC, we can scale \textit{Oz}MAC's frequency by the corresponding ratio to match bMAC's compute latency and throughput.
Table \ref{tab:freq2} provides TSMC N5 PPA for bMAC (0.5 GHz) and \textit{Oz}MAC at throughput-matching frequencies.
16x16 \textit{Oz}MAC incurs 4.6x higher latency and hence is not considered here.

For the same throughput, INT4 (4x4) and INT8 (8x8) designs deliver close to 30\% improvement in power/energy, while mixed precision designs (4x8 and 8x16) achieve even higher improvements in power/energy by up to 46\%. 
Note that \textit{Oz}MAC can potentially deliver even higher throughput than bMAC by leveraging the remaining headroom in power reduction (29\% to 46\%) to further increase the frequency. 

\textbf{Key Takeaway:} \textit{Oz}MAC maintains superiority in area, power and energy efficiency at frequencies ranging from 500~MHz to 1.5~GHz, and can leverage relative frequency scaling to achieve equal or higher throughput compared to bMAC without adversely affecting its power or energy efficiency.

\section{Conclusions}
\label{conclusion}

This paper presents rigorous industry standard evaluation of \textit{Oz}MAC, an updated re-implementation of previously proposed Bit-Pragmatic (PRA) MAC design that performs a series of simple shift-and-add operations. It accounts for only the `1' bits in input binary value (skipping the `0' bits) thus leveraging bit sparsity in DL workloads. The main goal of this work is to assess practical commercial implementation potential of dynamic bit sparsity exploitation through such zero skipping MAC designs. 
%
We demonstrate the presence of high bit sparsity in eight state-of-the-art DL benchmarks.
We implement a wide range of \textit{Oz}MAC designs using commercial design tools and latest TSMC N5 process node, and obtain PPA results across various data precisions and clock frequencies. \textit{Oz}MAC shows substantial improvements in all three metrics: area (up to 30\%), power (up to 80\%) and energy (up to 46\%) relative to conventional binary bMAC. Finally, we demonstrate the significant power reduction of \textit{Oz}MAC and how this can be leveraged to increase throughput by increasing frequency without compromising area and energy efficiency benefits. Future work will evaluate a large array of \textit{Oz}MAC units in an actual DLA at the system level. We believe all DLAs targeting low precision  inference should adopt \textit{Oz}MAC design.





\bibliographystyle{IEEEtran}
\bibliography{refs}

\end{document}